# Stability of Spatial Smoothness and Cluster-Size Threshold Estimates in FMRI using AFNI

Robert W Cox and Paul A Taylor
Scientific and Statistical Computing Core
National Institute of Mental Health
Bethesda MD 20892 USA
robertcox@mail.nih.gov

## Abstract

In a recent analysis of FMRI datasets [K Mueller et al, *Front Hum Neurosci* 11:345], the estimated spatial smoothness parameters and the statistical significance of clusters were found to depend strongly on the resampled voxel size (for the same data, over a range of 1 to 3 mm) in one popular FMRI analysis software package (SPM12). High sensitivity of thresholding results on such an arbitrary parameter as final spatial grid size is an undesirable feature in a processing pipeline. Here, we examine the stability of spatial smoothness and cluster-volume threshold estimates with respect to voxel resampling size in the AFNI software package's pipeline. A publicly available collection of resting-state and task FMRI datasets from 78 subjects was analyzed using standard processing steps in AFNI. We found that the spatial smoothness and cluster-volume thresholds are fairly stable over the voxel resampling size range of 1 to 3 mm, in contradistinction to the reported results from SPM12.

## Introduction

In 2016, a widely cited paper [1] was published, stating that the global false positive rate (FPR) in FMRI task activation mapping was highly inflated over the nominal 5% when using the cluster-thresholding methods implemented in several widely used software packages (AFNI, FSL, and SPM). One noteworthy cause for inflated FPRs common to these software tools was mis-estimation of the spatial smoothness of the noise in FMRI datasets, by using Gaussian-shaped approximations when the spatial autocorrelation actually tends to have heavier tails [1,2]. A recent follow-on commentary by Mueller et al (2017) [3] probed the variability of the spatial smoothness estimates and cluster statistical significances in FMRI analyses as the voxel resampling size was varied. Those authors applied SPM12 with synthetic task timing (as in [1]) to 47 resting-state FMRI datasets, and found that the estimated noise smoothness Full Width at Half Maximum (FWHM) parameter varied markedly (typically by ~1.6 mm) as the chosen voxel resampling size varied (isotropic voxels with edge lengths from 1 to 3 mm). To be precise, the highly scattered SPM12 FWHM estimates universally decreased as the voxel resampling size (Δ) decreased; see Fig 1. This non-intuitive result meant that the cluster-size thresholds (or cluster *p*-value calculations) would be adjusted downwards as well to compensate for voxel size, so that clusters of the same size in the results would appear to be more significant at finer resampling sizes. In [3], Mueller et al also pointed out that the default resampling size in SPM12 is 2 mm, but that the SPM developers' reply [4] to the issues raised in [1] used a 3 mm resampling size.

As various software tools implement different techniques and methodologies throughout a processing pipeline, the spatial sensitivity results from [3] cannot be assumed to apply to all FMRI analytical packages *a priori*. As developers of AFNI, after reading [3] we felt it was important to perform similar tests of the stability of the smoothness estimates in our software package. For the present study, we downloaded an FMRI dataset collection from the OpenfMRI website (https://openfmri.org/ accession number ds000030) [5], and datasets from 78 control subjects were extracted for analysis using AFNI. A dataset analysis approach similar to that described in [3] was implemented, and we describe the dependency of smoothness estimates and cluster-size thresholds on spatial resampling size Δ here.

Figure 1. SPM12 smoothness estimates from 47 resting-state FMRI datasets; lines connect the estimates from each subject at the different resampling sizes. This graph is from Fig 1C of [3], and is used by permission and by terms of the Creative Commons Attribution License (CC BY https://creativecommons.org/licenses/by/4.0/). The datasets used in the present work (cf. Figs 2 and 3) are distinct from the datasets used in [3].

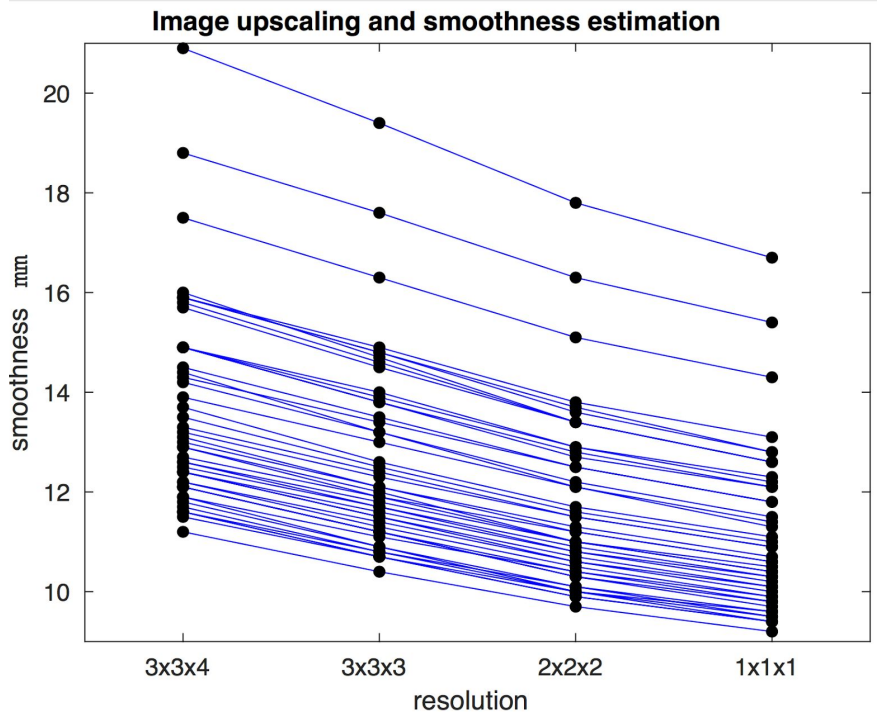

**Methods**

All analyses were implemented using AFNI v17.2.12 (Sep 2017), except for the SPM12 results generously provided by K Mueller from the work reported in [3]. (It is important to note that the analyses of [3] used a different collection of data than the analyses reported here.) Scripts of the AFNI commands used are provided in the Appendix, as well as the list of dataset IDs used in this study. For each subject, the resting-state and "pamenc" (pattern matching and encoding) task datasets were processed through the AFNI task-activation analysis pipeline, using the pamenc task timings provided with the datasets (in the case of the resting-state datasets, pseudo-task regression was intended to mimic the processing steps of [1,3]). The FMRI time series datasets were warped and resampled to MNI space at various voxel sizes (1, 2, and 3 mm), and 8 mm Gaussian FWHM blurring was applied (as in [3]) before task-based regression. The original spatial resolution of these axial EPI datasets was 3×3×4 mm$^3$ (as in [3]); TR was 2 s.

Using the regression residual time series datasets (two for each subject: rest and pamenc conditions), FWHM estimates were computed in AFNI using the mixed model method [2], which allows for a more general spatial autocorrelation function (ACF) than the commonly used Gaussian ACF model. In AFNI, the correlation ρ of noise samples that are distance $r$ apart is now modeled using the Gaussian-plus-exponential expression $\rho(r) = a \exp(-r^2/2b^2) + (1-a)\exp(-r/c)$, where $(a,b,c)$ are parameters estimated using a nonlinear least squares fit to the sample spatial correlations of the residuals, with constraints $(0 \leq a \leq 1, \ b > 0, \ c > 0)$. The FWHM estimate is then calculated from the parameters $(a,b,c)$ for each dataset. The justification for and applicability of this ACF model in FMRI is presented in [2]. (FWHM estimates for the pure Gaussian shape ACF model were also calculated, and the stability results for those values—not presented here—are very similar to the results for the FWHM values from the more general mixed model ACF.)

Given the $(a,b,c)$ parameters, a Monte Carlo simulation of Gaussian deviates with the same ACF was used to generate realizations of "pure noise" 3D datasets. These realizations are thresholded at various per-voxel *p*-values (*p*=0.001 in the results shown in Fig 3, as advised in [1,2]). The voxel count for the biggest cluster in each realization is saved into a table; 10000 3D realizations are used to find the cluster-size threshold that yields a 5% global FPR. That is, the cluster-size threshold is taken as the 500[th] largest value in the table of maximal cluster sizes. This brute force technique is how AFNI avoids the use of an asymptotic formula for cluster-size thresholds, and allows the ACF to be generalized from the Gaussian model.

# Results

The smoothness estimates are plotted vs $\Delta$ for each task case (pamenc and rest) in Figure 2a,b (compare with Fig 1 as taken from [3]; note the vertical scales are very different, with a range of 3 mm here and 12 mm in Fig 1C of [3]). The overall mean FWHM across all cases is 11.6 mm, with standard deviation 0.53 mm. Table 1 shows statistics of the changes in FWHM estimates from AFNI as $\Delta$ changes, with results from the pamenc and rest datasets shown separately. Although a majority of datasets analyzed with AFNI show increases in FWHM estimates as $\Delta$ decreases (the opposite result from SPM12 in [3]), there are a few (10/156) datasets where the FWHM estimates decrease as $\Delta$ decreases. The statistics of changes in smoothness estimates from the SPM12 results in [3] are shown to illustrate the qualitative difference from AFNI results.

Table 1. Changes in FWHM estimates $W(\Delta)$ as resampling size $\Delta$ shrinks (mean ± standard deviation). Paired *t*-tests between the AFNI pamenc and rest FWHM changes are all statistically insignificant (smallest *p*=0.52). SPM12 numerical results for data used for Fig 1C in [3] were kindly supplied by K Mueller (which were computed from a completely different set of data than used in this paper).

| $W(\Delta = 2) - W(\Delta = 3)$ | $W(\Delta = 1) - W(\Delta = 2)$ | $W(\Delta = 1) - W(\Delta = 3)$ |
|---|---|---|
| AFNI pamenc: +0.22 ± 0.18 mm | AFNI pamenc: +0.14 ± 0.22 mm | AFNI pamenc: +0.36 ± 0.23 mm |
| AFNI rest: +0.21 ± 0.17 mm | AFNI rest: +0.16 ± 0.20 mm | AFNI rest: +0.37 ± 0.23 mm |
| SPM12 rest: -0.96 ± 0.20 mm | SPM12 rest: -0.67 ± 0.13 mm | SPM rest: -1.62 ± 0.33 mm |

Cluster-size threshold volumes computed from the $(a, b, c)$ parameters for each case are plotted in Figure 3a,b (there is no directly analogous figure in [3]). The overall mean cluster-size threshold is 810 mm$^3$, with standard deviation 90 mm$^3$. Table 2 shows statistics of changes in cluster-size threshold estimates as $\Delta$ changes, with results from pamenc and rest datasets shown separately. (Note: these are cluster-size thresholds for *individual* datasets, not for group analyses.)

Table 2. Changes in cluster-size threshold estimates $C(\Delta)$ as resampling size $\Delta$ shrinks (mean ± standard deviation). Paired *t*-tests between the AFNI pamenc and rest cluster-size threshold changes are all statistically insignificant (smallest *p*=0.12).

| $C(\Delta = 2) - C(\Delta = 3)$ | $C(\Delta = 1) - C(\Delta = 2)$ | $C(\Delta = 1) - C(\Delta = 3)$ |
|---|---|---|
| AFNI pamenc: -3.3 ± 28.7 mm$^3$ | AFNI pamenc: +50.8 ± 33.2 mm$^3$ | AFNI pamenc: +47.5 ± 38.3 mm$^3$ |
| AFNI rest: +1.2 ± 26.0 mm$^3$ | AFNI rest: +54.2 ± 30.5 mm$^3$ | AFNI rest: +55.4 ± 35.4 mm$^3$ |

Figure 2a. AFNI smoothness estimates for the residuals dataset in the pamenc task for all 78 subjects at voxel resampling sizes of 3, 2, and 1 mm. Values for each subject are connected by dotted lines.

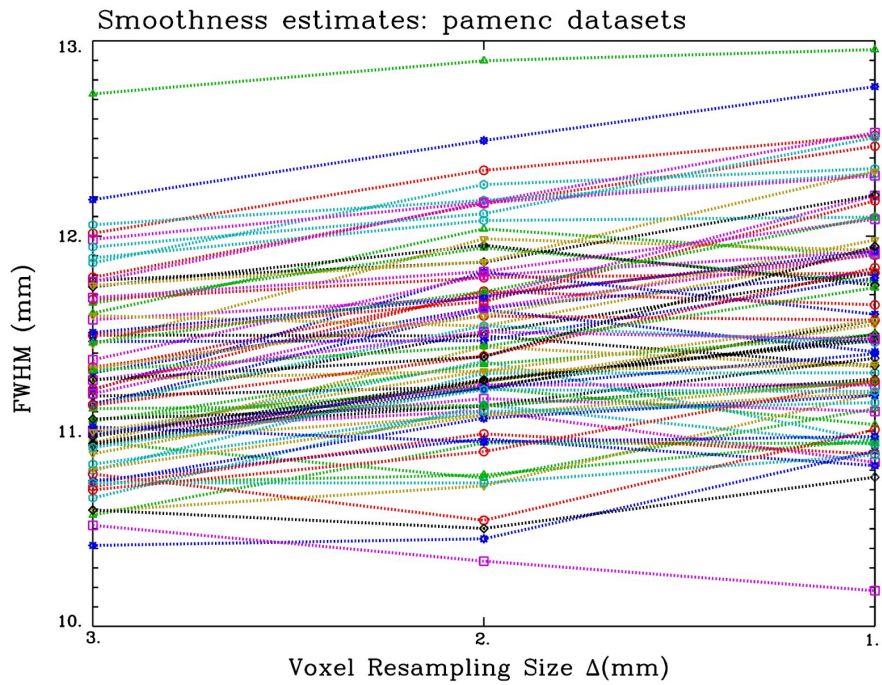

Figure 2b. AFNI smoothness estimates for the residuals dataset in the rest condition for all 78 subjects at voxel resampling sizes of 3, 2, and 1 mm.

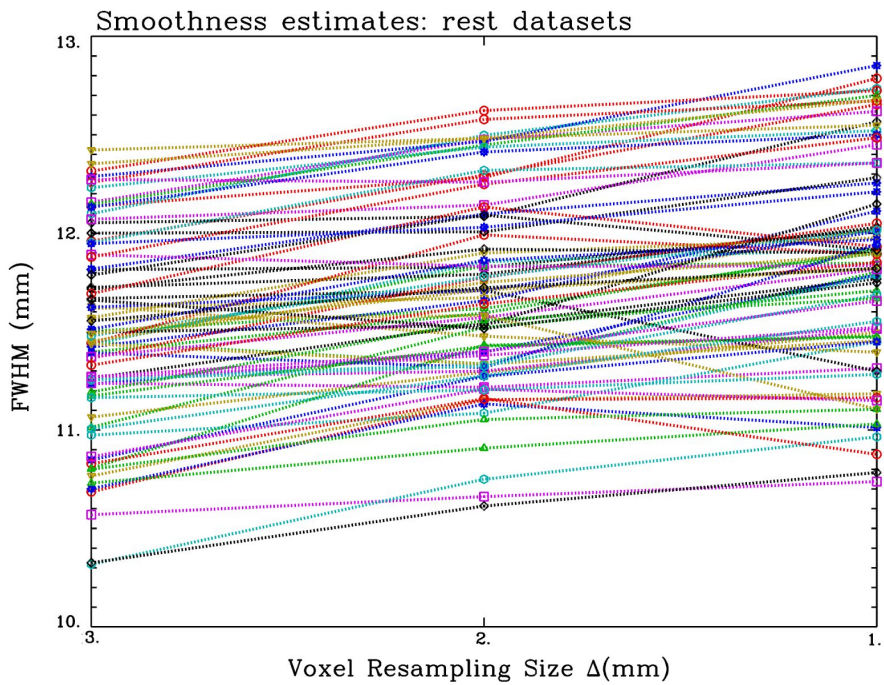

Figure 3a. AFNI cluster-size threshold estimates for the residuals dataset in the pamenc task for all 78 subjects at voxel resampling sizes of 3, 2, and 1 mm. The per-voxel threshold is set to *p*=0.001, and the cluster-size threshold is computed to provide a global false positive rate of 0.05. Values for each subject are connected by dotted lines.

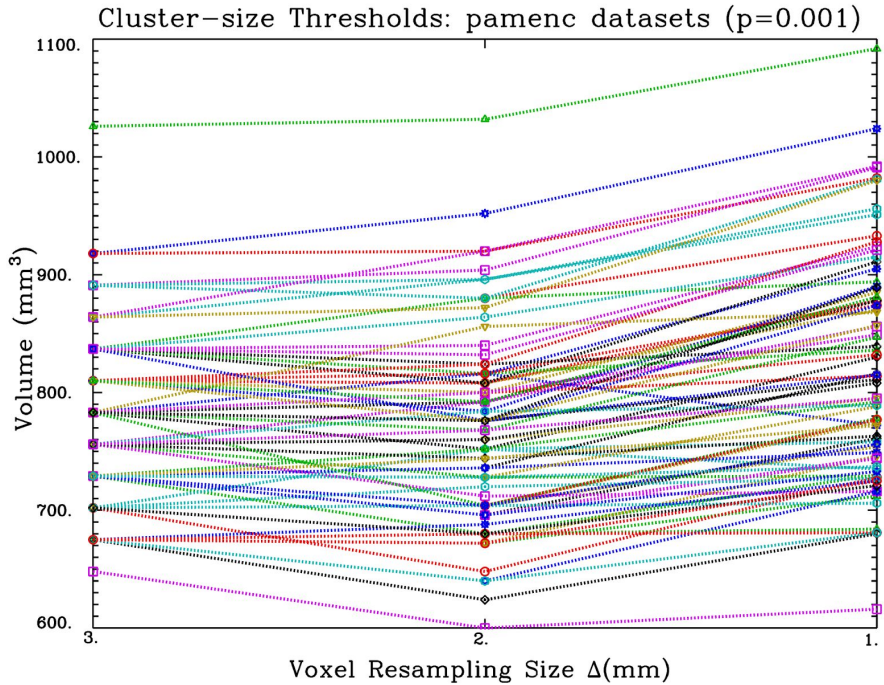

Figure 3b. AFNI cluster-size threshold estimates for the residuals dataset in the rest condition for all 78 subjects at voxel resampling sizes of 3, 2, and 1 mm.

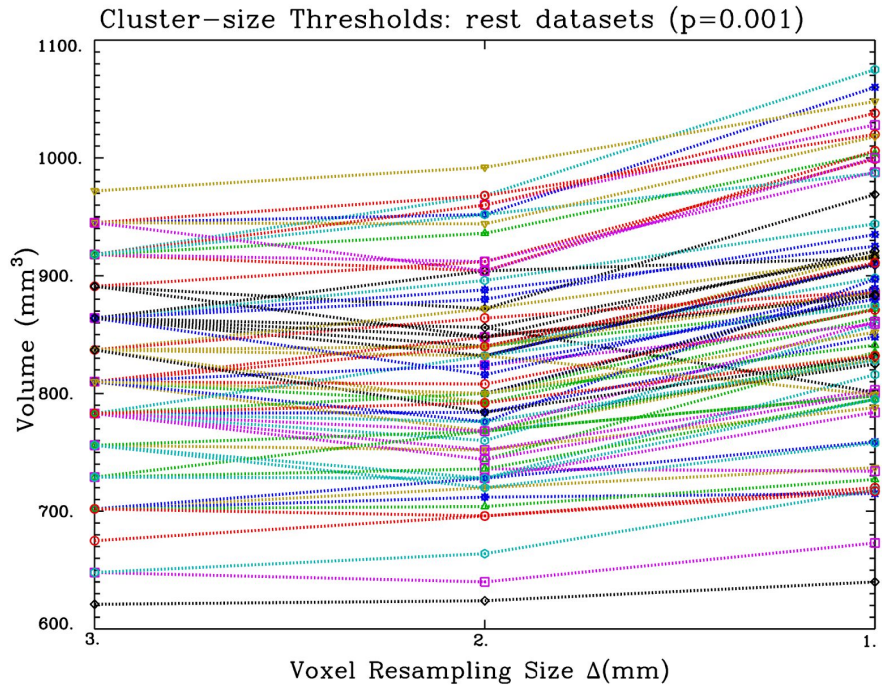

**Discussion**

Here, we have investigated the potential bias and dependence of FMRI clustering results on voxel resampling. Such dependence was reported previously using the SPM12 package [3], and as developers of the AFNI tools, we performed similar tests using a standard AFNI pipeline. The choice to change spatial resolution in a pipeline may be made, for example, in order to blur data slightly, to present results on a common grid used in other studies, or to present results with more continuous-looking overlays. However, upsampling cannot introduce new information into the data, and it can also greatly increase processing time[1]; the default pipeline in AFNI rounds voxel size to the nearest quarter mm (to leave data near their acquired spatial resolution).

Strong dependence of results on such a semi-arbitrary parameter as voxel size is highly undesirable and greatly reduces the robustness of reported results, which would depend both on the final grid size and very likely on the acquired spatial resolution, as well. Additionally, a further difficulty would be that datasets acquired from the same subject at different resolutions and then transformed to the same final voxel size might not have comparable spatial statistics simply due to processing issues, as different amounts of resampling would have been applied; this could affect longitudinal studies, as well as comparisons between studies acquired at different resolutions in general (e.g., meta-analyses). Possible sources of bias in software might include problems with the resampling technique itself, the method of smoothness estimation, the manner of cluster significance calculation, or the use of inappropriate approximations; however, the exact details would have to be investigated (and fixed, if an important bias is found) by the software developers themselves.

AFNI's smoothness estimate (FWHM) shows a modest dependence on the voxel resampling size $\Delta$, but neither as extreme nor as uniform as demonstrated in [3] for the SPM12 package. The smoothness estimates from SPM12 *decrease* by about 1.6 mm going from $\Delta$ = 3 mm to $\Delta$ = 1 mm, whereas in AFNI the smoothness estimates *increase* by about 0.4 mm—a factor which is about 4 times smaller, much less than any voxel size, and has the opposite sign. The FWHM estimates computed in AFNI from the rest and pamenc residuals do not differ significantly (e.g., a paired *t*-test cannot distinguish these samples).

---

[1] Increasing the isotropic spatial resolution from 3 mm to 2 mm increases the total number of voxels by a factor of 3.375, and increasing resolution from 2 mm to 1 mm increases the voxel count by another factor of 8. In the datasets analyzed here, the in-mask voxel counts went from about 80K to 260K to 2000K.

The tests of AFNI here were made on a publicly available collection [5], which is different than that used in [3] (from a previous study of theirs) with SPM12. However, as both collections of datasets were acquired in very standard manners, it is reasonable to compare some of their most general resulting properties. Central to the issue of robustness and reproducibility of FMRI results, the significant and uniform dependence of smoothness estimates and clusters on resampling, which appeared in SPM12's results [3], does not appear within the AFNI results. Thus, the results of [3] do not appear to be general properties of FMRI analyses, but instead depend strongly on the software tools and underlying techniques involved (in their case, SPM12). Strong dependence of smoothness and clustering on resampled voxel size is an undesirable trait in a software package; as noted above, there can be several steps within a processing pipeline within which voxel size bias may be introduced, and so any software demonstrating such dependence may need to test and update several features. The AFNI results reported herein do not appear have such a dependence, so that a change in this software's present resampling techniques and smoothness estimation do not appear necessary at this time.

Finally, in connecting smoothness estimates with final cluster results, the cluster-size threshold results in Fig 3 generally follow the trend of FWHM changes as $\Delta$ changes. Therefore, the global false positive rate in AFNI, using the mixed model ACF method described in [2], is not particularly sensitive to the voxel resampling size.

## Acknowledgements

The research for and writing of this paper were supported by the NIMH Intramural Research Programs (ZICMH002888) of the NIH/DHHS/USA. This work extensively utilized the computational resources of the NIH HPC Biowulf cluster (http://hpc.nih.gov). A plethora of thanks go to K Mueller for providing a hi-res copy of Fig 1C from [3] and the numerical values used to create that plot.

**Appendix: AFNI Processing Scripts**

All scripts were written to run on the NIH Biowulf cluster of Linux nodes, which is managed using the Slurm system (https://slurm.schedmd.com/). In a few locations, Slurm-specific variables are used to set the environment for processing. These scripts are written in the Unix shell language tcsh. AFNI itself can be downloaded in binary form by following the instructions at https://afni.nimh.nih.gov/pub/dist/doc/htmldoc/index.html or in source code form at https://github.com/afni/afni .

Each subject's data is stored in a subdirectory whose name is the subject ID (e.g., sub-10506). Within that directory, the task and resting state datasets from the OpenFMRI download are stored (e.g., sub-10506_task-pamenc_bold.nii.gz and sub-10506_task-rest_bold.nii.gz), as well as the task timings for the PAMENC and CONTROL tasks, and the nonlinear warp files to MNI space, which were computed separately using the AFNI @SSwarper script (supplied with AFNI source and binaries).

**List of subjects used:**
The 78 control (non-patient) subjects in the UCLA Phenomics study who had both "task-rest" and "task-pamenc" datasets to be analyzed [5]:
```
10506 10517 10523 10525 10527 10530 10557 10565 10570 10575
10624 10629 10631 10638 10668 10672 10674 10678 10680 10686
10692 10696 10697 10704 10707 10708 10719 10724 10746 10762
10779 10785 10788 10844 10855 10871 10877 10882 10891 10893
10912 10934 10940 10949 10958 10963 10968 10975 10977 10987
11019 11030 11044 11050 11052 11059 11061 11062 11066 11067
11068 11077 11088 11090 11097 11098 11104 11105 11106 11108
11112 11122 11128 11131 11142 11143 11149 11156
```

**Script A**: **Time series analysis for one subject**

The desired results of this script are the residuals for each input time series dataset ("task-pamenc" and "task-rest"—the file naming system in the download), at the given applied blurring radius (8 mm), and at the given voxel resampling size $\Delta$; 78×3 copies of Script A are submitted to the cluster (one for each subject and voxel resampling size combination). For example, all the AFNI processing results for "task-rest" with 8 mm blurring and resampling voxel size of 2 mm will end up stored in output directory sub-10506.b8mm.rest.R2.results. The relevant residuals dataset therein will be named errts.sub-10506_REML+tlrc, and it is this file which will be processed using Script B to produce one data point for each of Figures 2 and 3.

```
#!/bin/tcsh
#### Analyze one subject from the UCLA study,
#### for the pamenc stimulus and for the rest case.
# argv[1] = subject ID to run    [e.g., sub-12345]
# argv[2] = blur radius          [e.g., 4, 6, 8, 10]
# argv[3] = resampling size      [e.g., 3, 2, 1]

if( $#argv < 3 )then
  echo "Need 3 args: subj blur resam" ; exit 1
endif

set subj     = $argv[1]
set blur     = $argv[2]
set blurname = b${blur}mm
set resam    = $argv[3]

# topdir = where all the data lies buried
set topdir = /data/NIMH_SSCC/UCLA.pamenc/data_orig

# directory for this one subject's original data
set subdir = $topdir/$subj
if( ! -d $subdir )then
  echo "No $subdir -- exiting" ; exit 1
endif

# switch to the subject's data directory
cd $subdir

# uncompress any compressed datasets, for speed of I/O
set nzz = `find . -maxdepth 1 -name \*.nii.gz | wc -l`
if( $nzz > 0 ) gzip -d *.nii.gz
```

```
# pamenc stimuli info
# (extracted from ${subj}_task-pamenc_events.tsv in BIDS collection)
set stimfileC = pamenc.CONTROL.txt
set stimfileT = pamenc.TASK.txt
set stimnameC = CONTROL
set stimnameT = TASK
set stimresp  = dmBLOCK

# set control variables from SLURM
if( $?SLURM_CPUS_PER_TASK )then
 setenv OMP_NUM_THREADS $SLURM_CPUS_PER_TASK
endif
if( $?SLURM_JOB_ID )then
 set TEMPDIR = /lscratch/$SLURM_JOB_ID
endif

# MNI template supplied with AFNI
set tpath = `@FindAfniDsetPath MNI152_2009_template.nii.gz`
if( "$tpath" == "" )then
  echo "**ERROR: can't find template MNI152_2009_template.nii.gz" ; exit 1
endif
set bset = $tpath/MNI152_2009_template.nii.gz

# set some AFNI environment variables
setenv AFNI_COMPRESSOR     NONE
setenv AFNI_DONT_LOGFILE   YES

#### Process this one subject

set anat_dset = anatSS.${subj}.nii
if( ! -f $anat_dset )then
  echo "$anat_dset does not exist -- ERROR" ; exit 1
endif

## loop over pair of datasets
foreach task ( pamenc rest )

  set task_dset = ${subj}_task-${task}_bold.nii

# Check if FMRI results already exists
  set odir = ${subj}.$blurname.${task}.R${resam}.results
  if( ! -d $odir && -f $task_dset )then
    if( $?TEMPDIR )then
      set todir = $TEMPDIR/$task
    else
      set todir = $odir
    endif
```

```
# run afni_proc.py to create a single subject processing script

afni_proc.py -subj_id $subj                                     \
    -out_dir $todir                                             \
    -script  proc.$subj.$blurname.${task}.R${resam}             \
                     -scr_overwrite                             \
    -blocks despike tshift align tlrc volreg blur mask scale regress \
    -copy_anat $anat_dset                                       \
        -anat_has_skull no                                      \
    -dsets $task_dset                                           \
    -tcat_remove_first_trs 0                                    \
    -align_opts_aea -ginormous_move -deoblique on               \
         -cost lpc+ZZ                                           \
    -volreg_align_to MIN_OUTLIER                                \
    -volreg_align_e2a                                           \
    -volreg_tlrc_warp                                           \
    -tlrc_base $bset                                            \
    -tlrc_NL_warp                                               \
    -tlrc_NL_warped_dsets                                       \
          anatQQ.${subj}.nii                                    \
          anatQQ.aff12.1D                                       \
          anatQQ.${subj}_WARP.nii                               \
    -volreg_warp_dxyz $resam                                    \
    -blur_size $blur                                            \
    -blur_in_mask yes                                           \
    -regress_anaticor_fast                                      \
    -regress_anaticor_fwhm 20                                   \
    -regress_stim_times  $stimfileC $stimfileT                  \
    -regress_stim_labels $stimnameC $stimnameT                  \
    -regress_stim_types AM1                                     \
    -regress_basis "$stimresp"                                  \
    -regress_censor_motion 0.2                                  \
    -regress_censor_outliers 0.02                               \
    -regress_3dD_stop                                           \
    -regress_make_ideal_sum sum_ideal.1D                        \
    -regress_est_blur_errts                                     \
    -regress_reml_exec                                          \
    -regress_run_clustsim no

# Run analysis

echo "===== Starting analysis ====="
tcsh -xef proc.${subj}.$blurname.${task}.R${resam} \
     |& tee proc.${subj}.$blurname.${task}.R${resam}.output
```

```
      # If finished, clean up the results
      if( -d $todir )then
        pushd $todir
        # compress output dataset files
        gzip -1v *.BRIK *.nii
        popd
        # copy them to final resting space from temp space
        if( "$todir" != "$odir" )then
          mkdir -p $odir
          mv -prf $todir/* $odir/
        endif
      endif

   else
      echo "------------------------------------------------------------"
      echo "Skipping $subj for $task"
      if(   -d $odir     ) echo "  $odir EXISTS"
      if( ! -f $anat_dset ) echo "  $anat_dset DOES NOT EXIST"
      if( ! -f $task_dset ) echo "  $task_dset DOES NOT EXIST"
      echo "------------------------------------------------------------"
   endif

# end of loop over pair of datasets
end

# finished
exit 0
```

**Script B: Computing smoothness and cluster-size threshold estimates**

The following script is run for each subject, once Script A has finished successfully. It estimates the spatial smoothness parameter (FWHM in mm) for each condition (pamenc and rest), for each voxel resampling size (1, 2, and 3 mm). The estimation is done via AFNI program 3dFWHMx, using the (default) mixed model ACF method described in [2]. Analysis outputs from each subject (Script A) include a "full" mask of the high-intensity part of the EPI dataset, as transformed to MNI space; the computations in Script B are restricted to the voxels in this mask. The output comprises a set of text files containing the FWHM estimates and the cluster-size threshold estimates for various per-voxel *p*-value thresholds.

```
#!/bin/tcsh
if( $?SLURM_CPUS_PER_TASK )then
 setenv OMP_NUM_THREADS $SLURM_CPUS_PER_TASK
endif

# set subject from command line argument
set subj  = $argv[1]
# set blurring case: 8 mm
set blur  = b8mm
#choose clustering case: NN2 with 1-sided t-tests
set ccase = NN2_1sided.1D

set topdir = /data/NIMH_SSCC/UCLA.pamenc/data_orig/
cd   $topdir/$subj

# loops over datasets, and over resampling sizes
foreach task ( pamenc rest )
foreach resam ( R1 R2 R3 )

# check if results already exist
  if( -f blurs.$task.$resam.txt     && \
      -f csiz.ACFm.$task.$resam.txt     )then
    echo "=== skipping $subj $task $resam -- outputs extant"
    continue
  endif

# compute voxel volume
  set rvox = `echo $resam | sed -e 's/R//'`
  set vmul = `ccalc -form '%.2f' "$rvox^3"`
# enter AFNI output directory for this subject and case
  if( ! -d $subj.$blur.$task.$resam.results )then
    echo "=== no $subj.$blur.$task.$resam.results" ; continue
  endif
```

```
  pushd $subj.$blur.$task.$resam.results
  set mask = full_mask.${subj}+tlrc.HEAD
  if( ! -f $mask || ! -f errts.${subj}_REML+tlrc.HEAD )then
    echo "=== No $mask and/or No errts.${subj}_REML+tlrc.HEAD"
    popd ; continue
  endif

# estimate ACF parameters
  set blist = `3dFWHMx -mask $mask -acf NULL errts.${subj}_REML+tlrc`

# extract parameters for cluster-size threshold calculations
  set aaa  = $blist[5]
  set bbb  = $blist[6]
  set ccc  = $blist[7]
  set fff  = $blist[8]

# save FWHM to a file in the parent (subject) directory
  echo $fff > ../blurs.$task.$resam.txt

# cluster-size thresholds via mixed-model ACF
  3dClustSim -mask $mask -acf $aaa $bbb $ccc \
             -pthr 0.01 0.005 0.002 0.001    \
             -athr 0.05 -nodec               \
             -prefix Qsim
  set CsizACFm010 = `1dcat Qsim.$ccase'[1]{0}'`
  set CsizACFm005 = `1dcat Qsim.$ccase'[1]{1}'`
  set CsizACFm002 = `1dcat Qsim.$ccase'[1]{2}'`
  set CsizACFm001 = `1dcat Qsim.$ccase'[1]{3}'`
  \rm Qsim.*

# scale from voxel counts to microliters
  set q010 = `ccalc -form '%.2f' "$vmul*$CsizACFm010"`
  set q005 = `ccalc -form '%.2f' "$vmul*$CsizACFm005"`
  set q002 = `ccalc -form '%.2f' "$vmul*$CsizACFm002"`
  set q001 = `ccalc -form '%.2f' "$vmul*$CsizACFm001"`
  echo $q010 $q005 $q002 $q001 > ../csiz.ACFm.$task.$resam.txt

# back to the parent directory
  popd
end
end
exit 0
```

## Scripts C: Producing the Figures and Tables

Once all the results from 78 runs of Script B have been produced, the individual subject output text files are merged with a simple script:

```
#!/bin/tcsh
set topdir = /data/NIMH_SSCC/UCLA.pamenc/data_orig/
cd $topdir
foreach task ( pamenc rest )
foreach resam ( R1 R2 R3 )
  cat sub-*/blurs.$task.$resam.txt     > blurs.$task.$resam.txt
  cat sub-*/csiz.ACFm.$task.$resam.txt > csiz.ACFm.$task.$resam.txt
end
end
exit 0
```

The following script was used to produce the plots in Figs 2 and 3, using the (rather simple) AFNI plotting program 1dplot:

```
#!/bin/tcsh
# This script extracts data from the blurs.*.txt and csiz.*.txt
# files and makes some plots.

set topdir = /data/NIMH_SSCC/UCLA.pamenc/data_orig/
cd $topdir

##### Plots for blurring estimates (FWHM) vs resampling size #####

# set plot width (pixels)
setenv AFNI_1DPLOT_IMSIZE 2000

# reverse the order (3 2 1 instead of 1 2 3)
# requires manual surgery on the x-axis graph numbers :(
set dorev = 1

\rm blurs.*.png
foreach task ( pamenc rest )
  set bot = 10 ; set top = 13 ; set del =  3
  1dtranspose blurs.$task.R1.txt > q1.txt
  1dtranspose blurs.$task.R2.txt > q2.txt
  1dtranspose blurs.$task.R3.txt > q3.txt
  if( $dorev )then
    cat q3.txt q2.txt q1.txt > qq.txt
  else
    cat q1.txt q2.txt q3.txt > qq.txt
  endif
```

```
  1dplot -one -box -xzero 1 -dx 1                        \
         -yaxis ${bot}:${top}:${del}:10                  \
         -xaxis 1:3:2:1                                  \
         -dashed 3:3:3:3:3:3:3:3:3                       \
         -xlabel "Voxel Resampling Size \Delta\ (mm)"    \
         -ylabel "FWHM (mm)"                             \
         -plabel "Smoothness estimates: $task datasets"  \
         -png blurs.$task.ACFm.png                       \
         qq.txt
  \rm q?.txt
end

##### Plots for cluster-size thresholds (mm^3) vs resampling size #####

# Let's just do the p=0.001 voxel-wise threshold
set pval = 0.001 ; set pcol = 3
set cbot = ( 1000 600 700 )
set bot  =  600 ; set top  = 1100 ; set del  =  500
set maj  = 5
set meth = ACFm

# csiz.METH.TASK.RX.txt has 4 columns of cluster volumes

\rm csiz.*.png
foreach task ( pamenc rest )

  1dtranspose csiz.$meth.$task.R1.txt"[$pcol]" > q1.txt
  1dtranspose csiz.$meth.$task.R2.txt"[$pcol]" > q2.txt
  1dtranspose csiz.$meth.$task.R3.txt"[$pcol]" > q3.txt
  if( $dorev )then
    cat q3.txt q2.txt q1.txt > qq.txt
  else
    cat q1.txt q2.txt q3.txt > qq.txt
  endif
  1dplot -one -box -xzero 1 -dx 1                                \
         -yaxis ${bot}:${top}:${maj}:10                          \
         -xaxis 1:3:2:1                                          \
         -dashed 3:3:3:3:3:3:3:3:3                               \
         -xlabel "Voxel Resampling Size \Delta\ (mm)"            \
         -ylabel "Volume (mm^3)"                                 \
         -plabel "Cluster-size Thresholds: $task datasets (p=$pval)"  \
         -png csiz.$meth.$task.p=$pval.png                       \
         qq.txt
  \rm q?.txt
end
exit 0
```

The following script was used to produce the statistics used in Tables 1 and 2, using various AFNI command tools to calculate values and statistics from tabular files:

```
#!/bin/tcsh

set topdir = /data/NIMH_SSCC/UCLA.pamenc/data_orig/
cd $topdir

foreach blur ( R1 R2 R3 )
  1dcat blurs.pamenc.$blur.txt'[1]' > qp.txt
  1dcat blurs.rest.$blur.txt'[1]'   > qr.txt
  cat qp.txt qr.txt > blurs.bothACF.$blur.1D
  \rm qp.txt qr.txt
end

1deval -a blurs.bothACF.R3.1D -b blurs.bothACF.R2.1D -expr 'b-a' > qq.1D
set mm32 = ( `3dTstat -mean -stdevNOD -prefix stdout: qq.1D\'` )
echo "FHWM changes 3->2" $mm32
mv -f qq.1D q32.1D

set Pdif = ( `3dTstat -mean -stdevNOD -prefix stdout: q32.1D'{0..77}'\'` )
set Rdif = ( `3dTstat -mean -stdevNOD -prefix stdout: q32.1D'{78..$}'\'` )
set PRtt = ( `3dttest++ -no1sam -paired -setA q32.1D'{0..77}'\' \
                        -setB q32.1D'{78..$}'\' -prefix stdout:` )
echo "=== FWHM Stats for 3->2:"
echo "    pamenc mean diff = $Pdif[1] stdev = $Pdif[2]"
echo "    rest   mean diff = $Rdif[1] stdev = $Rdif[2]"
echo "    paired t-stat on pamenc-rest = $PRtt[2]"

1deval -a blurs.bothACF.R2.1D -b blurs.bothACF.R1.1D -expr 'b-a' > qq.1D
set mm21 = ( `3dTstat -mean -stdevNOD -prefix stdout: qq.1D\'` )
echo "FHWM changes 2->1" $mm21
mv -f qq.1D q21.1D

set Pdif = ( `3dTstat -mean -stdevNOD -prefix stdout: q21.1D'{0..77}'\'` )
set Rdif = ( `3dTstat -mean -stdevNOD -prefix stdout: q21.1D'{78..$}'\'` )
set PRtt = ( `3dttest++ -no1sam -paired -setA q21.1D'{0..77}'\' \
                        -setB q21.1D'{78..$}'\' -prefix stdout:` )
echo "=== FWHM Stats for 2->1:"
echo "    pamenc mean diff = $Pdif[1] stdev = $Pdif[2]"
echo "    rest   mean diff = $Rdif[1] stdev = $Rdif[2]"
echo "    paired t-stat on pamenc-rest = $PRtt[2]"

1deval -a blurs.bothACF.R3.1D -b blurs.bothACF.R1.1D -expr 'b-a' > qq.1D
set mm31 = ( `3dTstat -mean -stdevNOD -prefix stdout: qq.1D\'` )
echo "FHWM changes 3->1" $mm31
mv -f qq.1D q31.1D
```

```
set Pdif = ( `3dTstat -mean -stdevNOD -prefix stdout: q31.1D'{0..77}'\'` )
set Rdif = ( `3dTstat -mean -stdevNOD -prefix stdout: q31.1D'{78..$}'\'` )
set PRtt = ( `3dttest++ -no1sam -paired -setA q31.1D'{0..77}'\' \
                     -setB q31.1D'{78..$}'\' -prefix stdout:` )
echo "=== FWHM Stats for 3->1:"
echo "    pamenc mean diff = $Pdif[1] stdev = $Pdif[2]"
echo "    rest   mean diff = $Rdif[1] stdev = $Rdif[2]"
echo "    paired t-stat on pamenc-rest = $PRtt[2]"

foreach blur ( R1 R2 R3 )
  1dcat csiz.ACFm.pamenc.$blur.txt'[3]' > qp.txt
  1dcat csiz.ACFm.rest.$blur.txt'[3]'   > qr.txt
  cat qp.txt qr.txt > csiz.bothACF.$blur.1D
  \rm qp.txt qr.txt
end

1deval -a csiz.bothACF.R3.1D -b csiz.bothACF.R2.1D -expr 'b-a' > qq.1D
set mm32 = ( `3dTstat -mean -stdevNOD -prefix stdout: qq.1D\'` )
echo "Csize changes 3->2" $mm32
mv -f qq.1D r32.1D

set Pdif = ( `3dTstat -mean -stdevNOD -prefix stdout: r32.1D'{0..77}'\'` )
set Rdif = ( `3dTstat -mean -stdevNOD -prefix stdout: r32.1D'{78..$}'\'` )
set PRtt = ( `3dttest++ -no1sam -paired -setA r32.1D'{0..77}'\' -setB
r32.1D'{78..$}'\' -prefix stdout:` )
echo "=== Csize Stats for 3->2:"
echo "    pamenc mean diff = $Pdif[1] stdev = $Pdif[2]"
echo "    rest   mean diff = $Rdif[1] stdev = $Rdif[2]"
echo "    paired t-stat on pamenc-rest = $PRtt[2]"

1deval -a csiz.bothACF.R2.1D -b csiz.bothACF.R1.1D -expr 'b-a' > qq.1D
set mm21 = ( `3dTstat -mean -stdevNOD -prefix stdout: qq.1D\'` )
echo "Csize changes 2->1" $mm21
mv -f qq.1D r21.1D

set Pdif = ( `3dTstat -mean -stdevNOD -prefix stdout: r21.1D'{0..77}'\'` )
set Rdif = ( `3dTstat -mean -stdevNOD -prefix stdout: r21.1D'{78..$}'\'` )
set PRtt = ( `3dttest++ -no1sam -paired -setA r21.1D'{0..77}'\' -setB
r21.1D'{78..$}'\' -prefix stdout:` )
echo "=== Csize Stats for 2->1:"
echo "    pamenc mean diff = $Pdif[1] stdev = $Pdif[2]"
echo "    rest   mean diff = $Rdif[1] stdev = $Rdif[2]"
echo "    paired t-stat on pamenc-rest = $PRtt[2]"
```

```
1deval -a csiz.bothACF.R3.1D -b csiz.bothACF.R1.1D -expr 'b-a' > qq.1D
set mm31 = ( `3dTstat -mean -stdevNOD -prefix stdout: qq.1D\'` )
echo "Csize changes 3->1" $mm31
mv -f qq.1D r31.1D

set Pdif = ( `3dTstat -mean -stdevNOD -prefix stdout: r31.1D'{0..77}'\'` )
set Rdif = ( `3dTstat -mean -stdevNOD -prefix stdout: r31.1D'{78..$}'\'` )
set PRtt = ( `3dttest++ -no1sam -paired -setA r31.1D'{0..77}'\' -setB r31.1D'{78..$}'\' -prefix stdout:` )
echo "=== Csize Stats for 3->1:"
echo "    pamenc mean diff = $Pdif[1] stdev = $Pdif[2]"
echo "    rest   mean diff = $Rdif[1] stdev = $Rdif[2]"
echo "    paired t-stat on pamenc-rest = $PRtt[2]"
```